\documentclass[12pt,preprint]{aastex}
\usepackage{rotating}
\usepackage{threeparttable}
\usepackage{epstopdf}

\begin{document}

\title{Multiple shock structures in a radio selected cluster of galaxies }
\shorttitle{Multiple cluster shocks}
\author{S. Brown\altaffilmark{1,2}, J. Duesterhoeft\altaffilmark{3}, and L. Rudnick\altaffilmark{3}}

\altaffiltext{1}{CSIRO Astronomy \& Space Science;}
\altaffiltext{2}{Bolton Fellow}
\altaffiltext{3}{University of Minnesota, 116 Church 
Street SE, Minneapolis, MN  55455;  corresponding author: larry@astro.umn.edu}

\begin{abstract} We present a new radio-selected cluster of galaxies, 0217+70, using observations from the Very Large Array and archival optical and X-ray data.  The new cluster is one of only seven known that has candidate double peripheral radio relics, and the second one of those with a giant radio halo (GRH), as well. It also contains unusual diffuse radio filaments {\it interior} to the peripheral relics, and a clumpy, elongated X-ray structure.  All of these indicate a very actively evolving system, with ongoing accretion and merger activity, illuminating a network of shocks, such as those first seen in numerical simulations. The peripheral relics are most easily understood as outgoing spherical merger shocks with large variations in brightness along them, likely reflecting the inhomogeneities in the shocks' magnetic fields .  The interior filaments could be projections of substructures from the sheet-like peripheral shocks, or they might be separate structures due to multiple accretion events.  ROSAT images show large-scale diffuse X-ray emission coincident with the GRH, and additional patchy diffuse emission that suggests a recent merger event.  This uniquely rich set of radio shocks and halo offer the possibility, with deeper X-ray, optical and data higher resolution radio observations, of testing the models of how shocks and turbulence couple to the relativistic plasma.  0217+70 is also over-luminous in the radio compared to the empirical radio-X-ray correlation for clusters -- the third example of such a system. This new population of diffuse radio emission opens up the possibility of probing low-mass cluster mergers with upcoming deep radio continuum surveys. \end{abstract}


\section{ INTRODUCTION}
Clusters of galaxies are the most massive gravitationally bound systems in the Universe, hosting the majority of the observable matter in the current epoch in the form of an $\sim$10$^{7-8}$~K intra-cluster medium (ICM). The ICM is believed to be shock heated during structure formation.  Current radio observations reveal spectacular Mpc-scale synchrotron emission associated with the ICM of some massive clusters of galaxies, while other clusters of comparable mass show no sign of diffuse radio emission even at much lower luminosity limits (Venturi et al. 2007; Brunetti et al. 2009). The origins of the cosmic-ray (CR) leptons and large- scale $\mu$G magnetic fields themselves are still an open question, though a clear correlation with merging or dynamically accreting clusters has emerged (e.g., Brunetti et al. 2009; Cassano et al. 2010). This radio emission is an indicator of recent energy input into the ICM, whether through merger/accretion shocks and turbulence or AGN outflows, and its properties can be used to trace dynamical activity in low-density regions inaccessible at other wavelengths (Rudnick et al. 2009). Numerical simulations have begun to reveal a rich network of shock structures throughout the ICM (Miniati et al. 2001; Ryu et al. 2003; Pfrommer et al. 2008; Battaglia et al. 2009, Skillman et al. 2010; Vazza et al. 2010). Within a distance a few times the virial radius of massive clusters, low Mach number {\it internal} shocks (Ryu et al. 2003) play an important role transferring kinetic to thermal energy, as well as accelerating cosmic-ray electrons/positrons (CRe$^{\pm}$) and potentially thermodynamically important cosmic-ray protons (CRp). 

Radio relics, commonly defined as elongated radio emission not associated with the cluster center or an active cluster radio galaxy (Giovannini \& Feretti 2004), are found in $\sim$30 clusters. Peripheral radio relics, on the outskirts of some massive clusters are the brightest of the cluster shock structures. These are interpreted as outwardly propagating merger shocks that reach higher Mach numbers as they enter the low-density outskirts of the ICM.  ``Double" peripheral radio relics are rare, being found in only seven clusters to date (van Weeren et al. 2009). Clusters showing double relics include A3667 (Rottgering et al. 1997), A3376 (Bagchi et al. 2006), RXC J1314.4-2515 (Venturi et al. 2007), ZwCl~2341.1+0000 (van Weeren et al. 2009), A2345 and A1240 (Bonafede et al. 2009). RXC J1314.4-2515 is the only other cluster know to host a radio halo and double peripheral radio relics. Most recently, CIZA J2242.8+5301 has been shown to have double radio relics (van Weeren et al. 2010), one of which exhibited unambiguous signs of shock-acceleration of CR electrons. 

 Despite their prevalence in simulations, detecting ICM shocks in the denser central regions of clusters is confounded by their low contrast with the X-ray halo emission, their low Mach numbers and their shorter synchrotron lifetimes (Pfrommer et al. 2007; Skillman 2010). However, recent work (e.g. Markevitch et al. 2005: Abell 520) shows that even weak shocks (M$<\sim$2) can accelerate electrons in the ICM.

In this Letter we present VLA and archival X-ray and optical data of the radio selected cluster 0217+70, discovered in a blind survey for diffuse radio emission in the Westerbork Northern Sky Survey (WENSS: Rengelink et al. 1997; Delain et al. 2006; Rudnick et al. 2006). In section $\S$2 we present the radio observations and images, and $\S$3 details the archival optical and X-ray data on the cluster. $\S$4 outlines our analysis of the unique features found in 0217+70, and we summarize our key points in $\S$5. In this paper, we assume $H_{o}=70$, $\Omega_{\Lambda}=0.7$, $\Omega_{M}=0.3$, and the synchrotron spectral index $\alpha$ is defined by $I_{\nu} \sim \nu^{-\alpha}$. 

\section{OBSERVATIONS \& DATA REDUCTION}

\subsection{Observations}
Observations of 0217+70 were made with the NRAO Very Large Array (VLA) at L (1.4~GHz)and P (0.3~GHz) bands. L-band observations were taken in the D configuration for 3.8 hours, and P-band observations were taken in the C and D configurations for 3.3 and 3.5 hours respectively. All observations were made in two sidebands, IF1 and IF2. L-Band observations were made with IF1 at 1.4649 GHz and IF2 at 1.3851 GHz, each with a 48.8 MHz bandwidth. P-Band observations were made with IF1 at 327.5 MHz and IF2 at 321.5625 MHz, each with a bandwidth of 12.5 MHz. Only LL and RR correlations were made, so no information on linear polarization is available. The flux calibrators 3C48 and 3C286 were used during each observation. 

\subsection{Total Intensity Calibration and Imaging}
\begin{figure}
\begin{center}
\includegraphics[width=14cm]{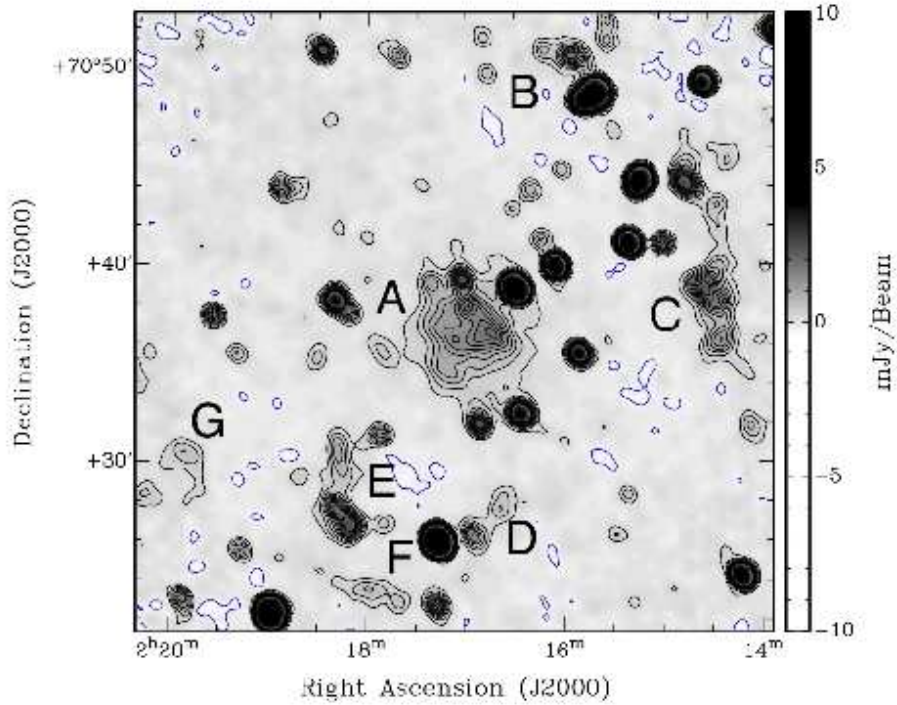}
\end{center}
\caption{\label{lband} VLA L-Band image of the 0217+70 source with contours starting at 2.5$\sigma$=210~$\mu$Jy~beam$^{-1}$ and increasing by 2.5$\sigma$ (blue is -2.5$\sigma$). Beam $= 49.52''\times43.10''$ at $30.03^{\circ}$. RMS noise is 86~$\mu$Jy~beam$^{-1}$. Labels are diffuse structures (see Fig. \ref{diff}).}
\end{figure}

\begin{figure}
\begin{center}
\includegraphics[width=14cm]{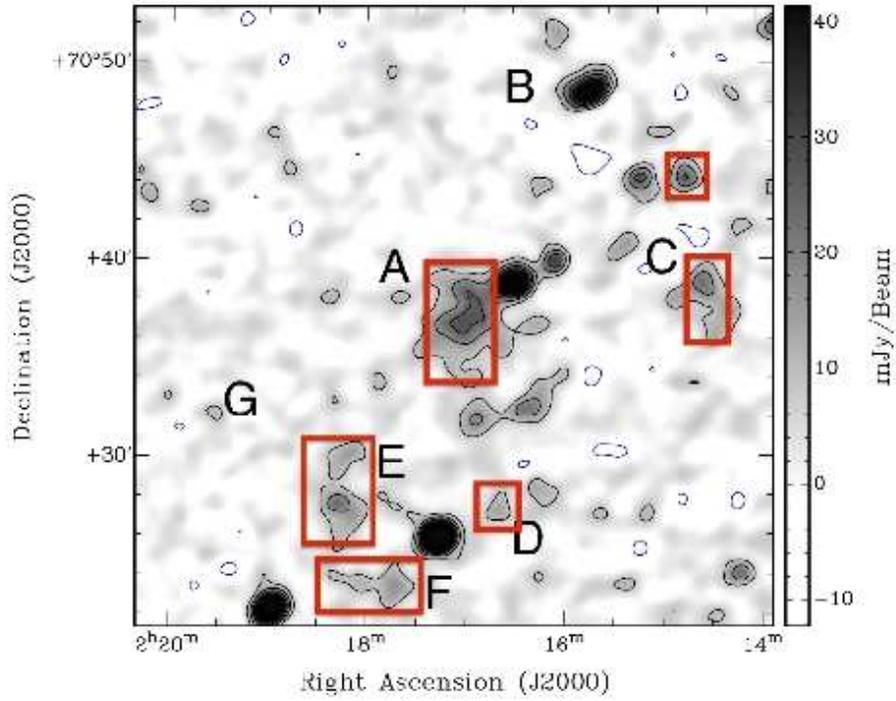}
\end{center}
\caption{\label{pband} VLA P-Band image of the 0217+70 source with contours 2.5$\sigma$=8.5~mJy~beam$^{-1}$ and increasing by 2.5$\sigma$  (blue is -2.5$\sigma$). Beam $=68.85''\times60.95''$ at $-62.78^{\circ}$. RMS noise is 3.4 mJy/beam. Red boxes are regions used to compute average spectral indices (see text).}
\end{figure}

The calibration and reduction of the VLA data was performed using the NRAO's Astronomical Image Processing System (AIPS). The VLA data were calibrated using standard procedures.  Images were created with the IMAGR task.  Several iterations of phase-only self-calibration were run on each data set using CALIB, finishing with a single round of amplitude self-calibration. IF1 and IF2 were calibrated separately for each observation.  Multiple IF1 and IF2 datasets resulting from observations in different configurations were then combined into single UV datasets using DBCON. Each image was cleaned using IMAGR with 9,000 clean components and a gain of 0.1, which was sufficient to clean slightly into the noise. IF1 and IF2 maps were combined with COMB using an rms noise-weighted average. A primary beam correction was applied to each final image using PBCOR. The P-Band image was interpolated to the geometry of the L-Band image using OHGEO. Beam and noise information are given in the figure captions.  VLA L-Band and P-Band images of 0217+70 are shown in Figures \ref{lband} and \ref{pband}, respectively. The L-band (P-band) flux of the central diffuse source A (Fig. \ref{lband} \& \ref{pband}) is 58.6$\pm$0.9 (326$\pm$30)~mJy (not including imbedded point sources), and its luminosity at z=0.0655 (see $\S$3.1) is 5.53 $\pm$0.09 (34.2$\pm$2.9)$\times 10^{23}~W~Hz^{-1}$. 

\begin{figure}
\begin{center}
\includegraphics[width=14cm]{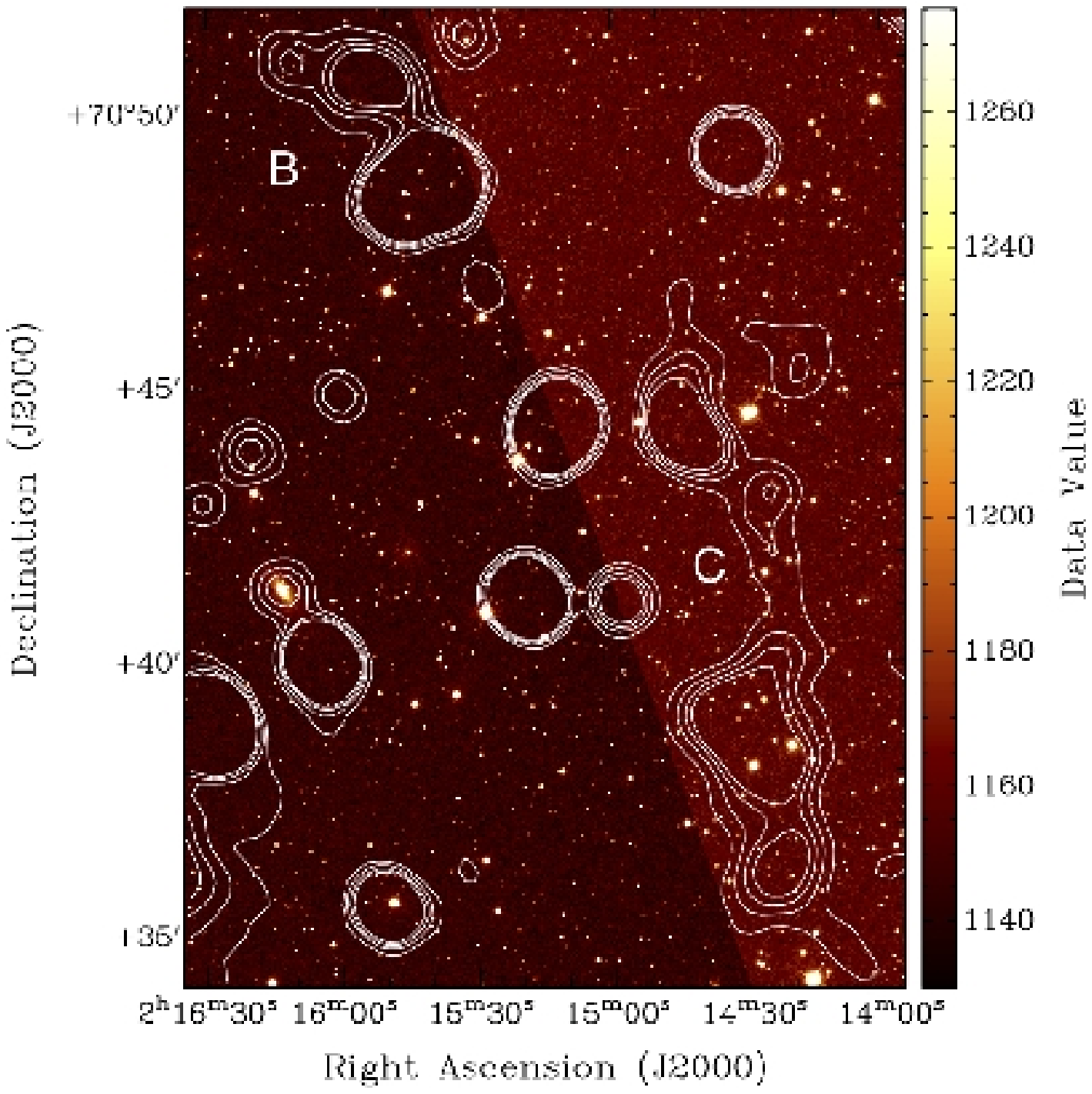}
\includegraphics[width=14cm]{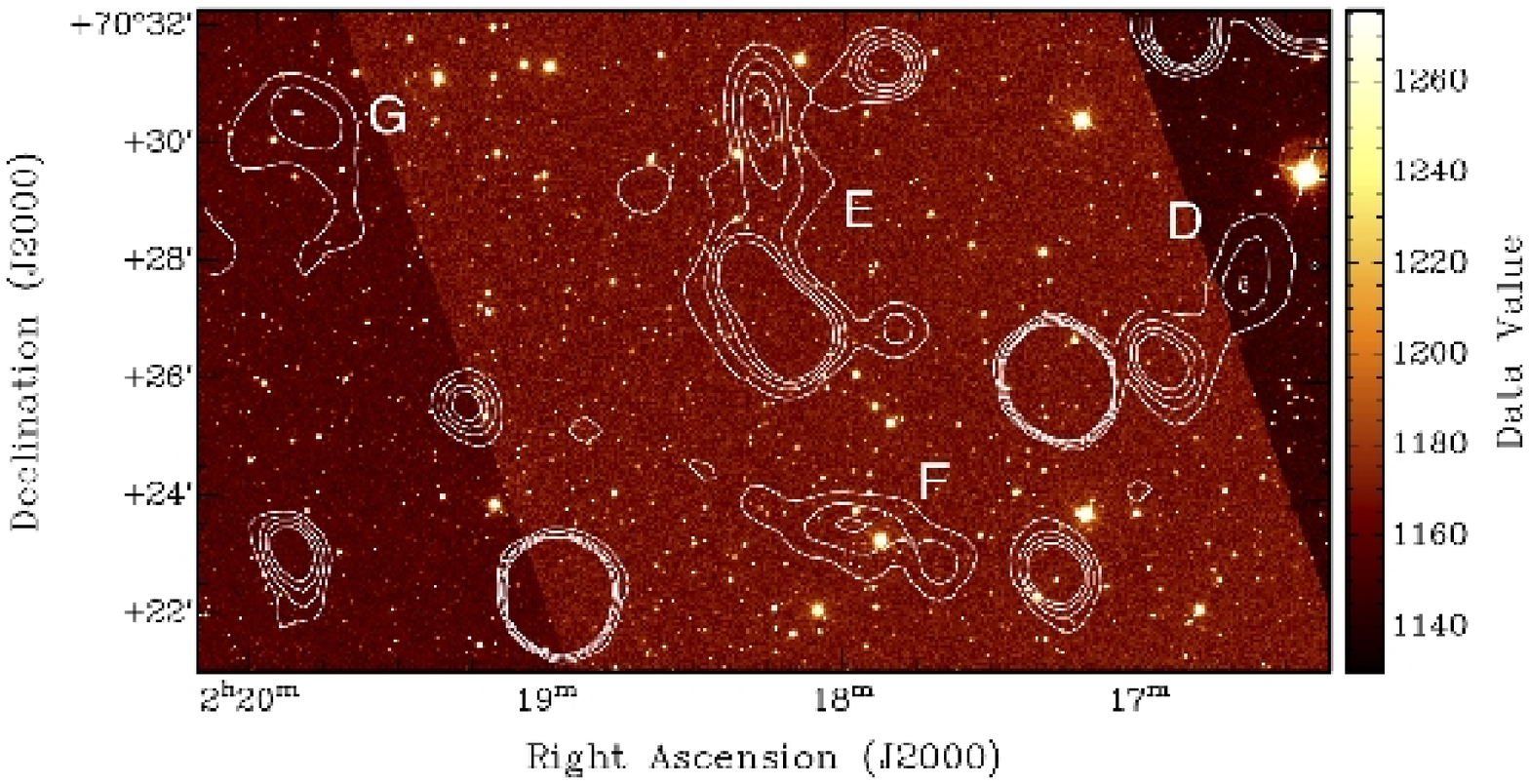}
\end{center}
\caption{\label{optical} SDSS R grayscale with VLA L-Band contours (same contours and resolution as Fig. \ref{lband}). Top: The Northern relics; Bottom: The Southern relics.}
\end{figure}

\subsection{Spectral Index}
A spectral index ($\alpha$) map of 0217+70 was created in AIPS using COMB. Pixels that were less than 2$\sigma$ above the background noise in each map were blanked. The sensitivity of the P-band observations were insufficient to detect  all of the diffuse emission seen in Fig. \ref{lband}. Average indices over the diffuse patches are A:~1.34$\pm$0.19; C:~1.34$\pm$0.16 (North) and 1.48$\pm$0.42 (South); D:~1.83$\pm$0.24, E:~1.52$\pm$0.38; F:~2.0$\pm$0.3.  These regions are indicated in Fig. \ref{pband} by red boxes. For B and G there was an insufficient amount of flux in the P-band image to calculate a spectral index. 

\section{Optical \& X-ray Identification}

\subsection{Optical}
The 0217+70 field is at the low galactic latitude of 9$^{\circ}$, with an estimated reddening of A$_V$=4.5, making the optical characterization difficult. A visual inspection of SDSS photometric redshifts suggested a possible clustering of at redshift z$\sim$0.065 within the central diffuse radio source. This excess was confirmed by comparing the surface density of redshift-binned galaxies, $n$, in a 400''$\times$400'' box centered on 0217+70  (in galaxies per arcsec$^{2}$) with the surface density in a larger 2$^{\circ}\times2^{\circ}$ box. This quantitative comparison also suggests a possible clustering of galaxies at redshift z$\sim$0.0655 within the central diffuse source of 0217+70.  However, the total excess number of galaxies with that redshift is small, with $n_{cluster}/n_{background}=6.55\pm2.09$, so deeper optical/infrared spectroscopy is needed. We also searched for excess galaxies around the other diffuse radio sources in the northwest and southeast, but no other significant concentrations at any redshift were observed. Fig. \ref{optical} shows an SDSS R band image with VLA L-Band contours for the northern and southern radio filaments.

\subsection{X-ray}
Diffuse X-ray emission in the ROSAT broadband images is seen coincident with the central diffuse radio emission and extending far beyond it in the northwest. We used webPIMMS\footnote{http://heasarc.nasa.gov/Tools/w3pimms.html} to convert the ROSAT  0.1-2.4 keV count rates into flux in erg s$^{-1}$ using a Raymond-Smith model with a plasma temperature of 1 keV as in Ponman et al. (1996). Galactic n$_{\mathrm{H}}$ values were determined by averaging the weighted average n$_{\mathrm{H}}$ values from both Dickey \& Lockman (1990) and Kalberla et al. (2005). The total flux (including the halo, bridge, and northwest source, see below) is 625$\times 10^{-14} erg~s^{-1}~cm^{2}$.

\begin{figure*}
\begin{center}
\includegraphics[width=14cm]{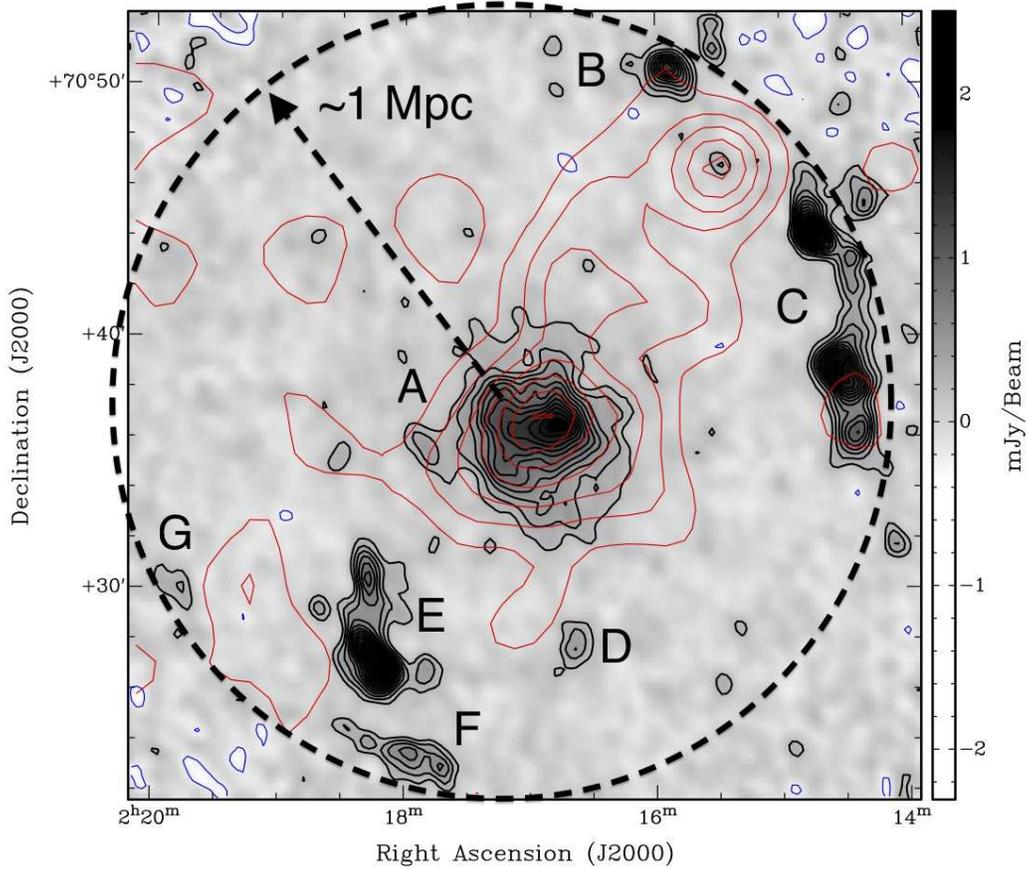}
\end{center}
\caption{\label{diff} Contours VLA L-Band image of the 0217+70 source with contours starting at 3$\sigma$=210~$\mu$Jy~beam$^{-1}$, increasing in steps of 2$\sigma$  (blue is -3$\sigma$). Beam $= 49.52''\times43.10''$ at $30.03^{\circ}$. RMS noise is 70~$\mu$Jy~beam$^{-1}$. Red contours are the Rosat All Sky Survey image (0.1-2.4 keV) convolved with a 3$^{\prime}$ gaussian kernel, in counts, starting at 5$\sigma$=7.5 and increasing in 2$\sigma$ intervals. The circle (and arrow base) is centered on the relics and is offset from the X-ray(radio) center by 90$^{\prime\prime}$(120$^{\prime\prime}$) or 113(150) kpc).}
\end{figure*}

Figure \ref{diff} shows the ROSAT broadband continuum emission in the region of 0217+70 with VLA 1.4~GHz contours showing diffuse emission. The diffuse radio emission was isolated by subtracting the point-sources out of the 1.4~GHz uv-data, and then re-mapping and cleaning.  The central X-ray emission covers the entire central diffuse radio source, and is identified in the ROSAT Bright Source Catalogue (Vogues et al. (1999)) as 1RXS J021649.0+703552. The bright, possibly unresolved X-ray peak in the northwest is unidentified and there are no obvious optical associations. There is very faint radio emission coincident with the NW X-ray peak; its origin is unknown. Connecting the central emission to the NW peak is a diffuse bridge of X-ray emission, which shows signs of substructure as well. We therefore assume that the bridge and NW unresolved source are related to the central X-ray source. The total X-ray luminosity of 0217+70 assuming z=0.0655 is then L$_{X}$=43.8 (log(erg s$^{-1}$)), which is of the same order of magnitude as  X-ray bright clusters (Ebeling et al. (1998)). The luminosity of the central diffuse halo alone is L$_{X}$=43.4 (log(erg s$^{-1}$)). 

\section{Discussion} 

\noindent {\sc Identification \& Classification}: We classify the central diffuse radio emission as a radio halo based on the optical excess and the diffuse X-ray emission with the same location and approximate size.  The total extent of the halo, when smoothed to a resolution of 80'' is $\sim$850~kpc, putting it in the ``giant radio halo'' category (R$_H>$300~kpc, e.g., Cassano et al. 2008). The lack of an apparent rich optical cluster is likely due to the large extinction. In addition, the elongated diffuse radio emission located on the peripheries with their major axes being nearly perpendicular with the core structure of a radio halo are what is expected for giant radio relics (Ferrari et al. 2008). We also observe steep spectral indices for the central and peripheral sources, which is typical of radio halos and relics.

0217+70 does not fit well with the well-established relation between radio halo luminosity and X-ray luminosity (Liang et al. 2000; Cassano et al. 2006; Brunetti et al. 2009); its radio luminosity is an order of magnitude above the extrapolation of this relation down to the low observed X-ray luminosity.
Another similar outlier is the small asymmetric halo in Abell 1213 (Giovannini et al. 2009), associated with a very low luminosity X-ray cluster.   A third related example is the low-mass radio relic system 0809+39 (Brown \& Rudnick 2009), whose radio emission is two orders of magnitude above a similar correlation for radio relics. The relatively high radio emission in these cases might  represent a new cluster population where the efficiency of energizing the relativistic plasma is especially high, or a special time during the merger process, where the radio luminosity peaks (Skillman et al. 2010).  In either case, radio selection of merging clusters may be possible with upcoming deep radio continuum surveys, e.g. the Evolutionary Map of the Universe (EMU) (Norris 2009) survey with the Australian Square Kilometre Array Pathfinder (ASKAP)\footnote{http://www.atnf.csiro.au/projects/askap/}, and the Surveys Key Science Project with Low Frequency Array (LOFAR)\footnote{http://www.lofar.org/}.

There are two factors which could have  influenced the observed radio/X-ray luminosity of 0217+70, and change the above conclusions.  Given the tentative nature of the redshift determination of 0.065, if it were as high as 0.2 then the cluster would fall on the upper end of the halo correlation, instead of the radio being far above the correlation.  However, at z=0.2 the halo would be more than 2 Mpc in total size, which would make it one of the largest halos ever found. The other uncertainty has to do with the hydrogen column density, which might be patchy at these low galactic latitudes.  Further X-ray observations are required to measure the absorption directly;  if it were three times larger than the nominal n$_H$=2.88E+21 used here (e.g. Dickey and Lockman 1990), then the inferred X-ray luminosity would increase enough to agree with the observed radio/X-ray correlation.\\

\noindent {\sc Peripheral Relics}: Fig. \ref{diff} shows the diffuse radio emission after point-source subtraction. Four  diffuse sources are found in the cluster outskirts, all at 1~Mpc distance from the cluster center. This correspondence suggests that these sources are related, perhaps being caused by two spherical merger shocks at a common distance from the cluster core. Higher resolution radio observations are needed to fully rule out extended radio galaxy origins for these sources, though the steep spectral indices support their identification as diffuse relics. Assuming this identification of the diffuse filaments B,C (northwest) and G,F (southeast) with symmetrical outgoing merger shocks, then we need to understand why these shocks are so irregularly illuminated, yielding prominent gaps. It is uncommon to see multiple significant peaks in observed radio relics, which are mostly uniform brightness filaments (Giovannini \& Feretti 2004).  Multiple emission regions in Abell 548 have been claimed to be part of a single shock structure (Solovyeva et al. 2008), though they are not as clearly aligned as those shown in Fig. \ref{diff}. 

Numerical simulations using 3D MHD show that these outgoing shocks are quite complex structures, with most of the derived synchrotron emission coming from filamentary and clumpy structures within an overall sheet-like geometry (E. Hallman, S. Skillman and H. Xu, 2010, private communication).   Sheet-like structures will be easiest to detect edge on, even if they have a low filling factor for strong-field, synchrotron emitting regions.  They will tend to be of more uniform brightness  in this projection than they would appear when face-on.   It is not yet clear how the strong field regions are related to the very local properties of the X-ray emitting plasma that dominates the mass and dynamics.  The gaps observed in 0217+70 and other peripheral relics could be related to the MHD behavior, or they could simply represent ``weather" in the evolving B field that is not well correlated to the overall density or temperature in the X-ray plasma.  Another possibility is that the gaps are a radiative phenomenon, e.g. due to increased losses in the higher fields that accompany higher thermal plasma densities (e.g., Miniati et al. 2001, Pfrommer et al. 2008, Skillman et al. 2010).  Pre-existing populations of relativistic particles, e.g, from AGNs, could also play a role in the brightness variations. Deep, low-frequency radio observations should reveal continuous synchrotron emission across the gaps if this scenario is correct.  A curious feature seen in Fig. \ref{diff} is that the ``gaps" are partially filled with X-ray emission, namely the bright secondary X-ray component in the northwest, and the diffuse 5$\sigma$ contour in the southeast; more sensitive X-ray observations are needed in order to map the extended thermal ICM in these regions.  \\

\noindent {\sc Internal Radio Filaments}: The diffuse sources D and E are also apparently unrelated to extended radio galaxies (Fig. \ref{optical}) and have elongated filamentary morphologies. Multiple (greater than 2) radio filaments have also been found at the edges of the halo in Abell 2255 (Pizzo et al. 2008; Pizzo \& de Bruyn 2009; Pizzo et al. 2010). Cosmological simulations that track CR acceleration during large-scale structure formation show multiple radio emitting shocks forming during major merger events for a wide range of cluster masses (Battaglia et al. 2009; Skillman et al. 2010).
Skillman et al. (2010) trace the temporal evolution of this radio emission which shows numerous shock structures persisting throughout the common X-ray envelope of the merging systems, in some cases for 2~Gyr.  

Another possible scenario is that the radio filaments are the result of continued infall onto the cluster, as opposed to being generated in a single merger event. Spectroscopic analysis looking for velocity substructures, as well as a deeper X-ray surface-brightness and temperature mapping, could reveal multiple merger events in 0217+70.  Finally, these apparently interior filaments may be projected from the outgoing merger shocks, if they are not being seen edge-on, as suggested by Pizzo et al. 2010) for Abell 2255 from the polarization characteristics of the filaments seen near the edge of the halo.  In any case, determining the geometry and then origin of such  shocks is critical to understanding the role they play in utilizing the infall energy to thermalize the ICM (Ryu et al. 2003). The next generation of deep cluster surveys will be especially useful in this area.  \\

\noindent {\sc Halo structure}: Another interesting feature of 0217+70 is that the northwestern (NW) edge of the halo shows a steeper cut-off than the rest of the halo, and is perpendicular to the merging direction. The NW edge is also adjacent to the diffuse extension of the X-rays that connect the halo to the NW peripheral relic(s) B \& C. A similar structure was reported in the Coma cluster (Brown \& Rudnick 2010), where a shock is found on the western edge of the radio/X-ray halo, connecting
to the diffuse X-ray bridge in the infall region to the southwest. Higher resolution X-ray imaging and temperature mapping, along with higher resolution radio observations, are needed in order to see whether the edge in 0217+70 corresponds to a shock-front. There is also an offset of $\sim$30$^{\prime\prime}$ between the diffuse radio and X-ray central halo peaks. Similar offsets have been observed on other clusters, including the Coma cluster (Deiss et al 1997), and are likely a result of the non-equilibrium state of the merging cluster system. 

Spectral indices, X-ray data, and our geometric model of the emitting regions are not of sufficient quality to estimate the Mach number of the shocks assumed to accelerate the synchrotron emitting CRe$^{\pm}$, though deeper observations in both radio and X-ray regimes could make 0217+70 an important test-bed for ICM physics. \\

\section{Summary} We have reported the detection of a new radio-selected merging cluster of galaxies hosting a giant radio halo, a rare candidate double peripheral relic system, enclosing at least two other diffuse radio filaments unassociated with AGN. ROSAT X-ray data shows emission coincident with the giant radio halo, a second, perhaps unresolved peak and a bridge of emission connecting these two, suggesting a recent merging event. The GRH is slightly offset from the X-ray peak and shows a sharper gradient/drop-off on the side adjacent to the bridge and sub-cluster. Both of these are further indications of merger dynamics. We are thus beginning to observe more of the shock network in clusters, as predicted in numerical simulations of cluster mergers (Battaglia et al. 2009; Skillman et al. 2010).  The fact that 0217+70 was selected on radio properties alone, as well as its deviation from the radio/X-ray luminosity correlation for GRHs, means that this system (along with 0809+39 and Abell 1213) may well be the beginning of a larger population of cluster radio sources that will be probed by upcoming sensitive all-sky radio surveys.      

We gratefully acknowledge help and advice from E. Hallman, T. W. Jones, and S. Skillman. 

Partial support for this work at the University of Minnesota comes from the U.S. National Science Foundation grant AST~0908688.  We acknowledge the use of NASA's SkyView facility located at NASA Goddard Space Flight Center. The Very Large Array is operated by the National Radio Astronomy Observatory (NRAO).  The NRAO is a facility of the National Science Foundation operated under cooperative agreement by Associated Universities, Inc. Archival observations obtained from XMM-Newton, an ESA science mission with instruments and contributions directly funded by ESA Member States and NASA, and ROSAT archives from HEASARC.

\end{document}